\documentclass[lettersize,Conference]{IEEEtran}
\usepackage[dvipsnames]{xcolor}
\usepackage{amsmath,amsfonts}
\usepackage{algorithmic}
\usepackage{algorithm}
\usepackage{array}
\usepackage[caption=false,font=normalsize,labelfont=sf,textfont=sf]{subfig}
\usepackage{textcomp}
\usepackage{stfloats}
\usepackage{url}
\usepackage{verbatim}
\usepackage{graphicx}
\usepackage{cite}
\usepackage{fancyhdr}
\newcommand{\Be}{\mathcal{B}\rm{e}}
\newcommand{\todo}{{\bf \textcolor{red}{TODO}}}

\begin{document}

\title{Phasor-Pursuit Directional Modulation}

\author{David Couto~\IEEEmembership{Member, IEEE}, Arash Samani~\IEEEmembership{Senior Member, IEEE}, Alec Yonika \\ BAE Systems Inc., FAST Labs \\ Merrimack, NH 03054}


\IEEEpubid{979-8-3503-9214-2/24/\$31.00~\copyright~2024 IEEE}

\maketitle

\begin{abstract}
Emitting phased array RF systems have to contend with an ever-increasing number of eavesdroppers as technological advancements provide lower cost and/or more capable radios. Often, eavesdroppers can accumulate sufficient information transmitted in sidelobes by integrating over long enough periods. Directional modulation (DM) disrupts this capability by inducing a randomized walk through IQ-space to reach a pertinent location which corresponds to a symbol of particular amplitude and phase. This results in higher secrecy capacity. The path taken by the cumulative element contributions are determined by the complex weights of individual transmitters. Because large phased arrays support a large number of available paths, repetitions of applied weights are not concerning. The same cannot be said for arrays that consist of only a few elements, e.g. WiFi routers. We introduce a computationally efficient and flexible framework for DM that utilizes real-valued phase rotations of element weights. It supports a wide family of modulation schemes in phase and/or amplitude. By employing the state-of-the-practice orthogonal noise injection framework, we demonstrate a richness of unique paths which resolves concerns about repeated weights. In our proposed scheme, there is a small reduction in received power compared to traditional beamforming, as little as 1 dB, which is an advantage over conventional directional modulation which typically sacrifices 6 dB of power. Also, there is a significantly larger set of possible element weights than that of the conventional scheme. This feature protects against the possibility of eavesdroppers breaking the distortion-based obfuscation of symbols over repeated observations. From these two key benefits, Phasor Pursuit Directional Modulation provides more secrecy capacity than conventional directional modulation via increased power delivery and increased receiver SNR and does so with resilience to advanced eavesdropping threats.
\end{abstract}


\begin{section}{Introduction}
\IEEEPARstart{F}{ielded} transmitting RF array systems require the means to protect their transmissions from potential eavesdroppers. One tactic is to protect the communication at the symbol level, one example of a technology in this family of solutions is directional modulation (DM). To provide security, DM relies upon obfuscation of the symbols in power radiated off-target, i.e. side lobes, if we consider the context of a traditional uniform linear array. This is done by augmenting transmitted signals at the element level such that the a-priori secure communication direction is the only direction where the waves combine with the intended phase and amplitude, which makes eavesdropping infeasible~\cite{DingFusco2014}. Contrasting with standard cryptographic approaches to protect transmitted content~\cite{Stallings}, DM can be implemented at the physical layer~\cite{DalyBernhard1,DalyBernhard2} for information theoretic security~\cite{BlochBarros}. DM has been described as either a direct manipulation of the antenna radiating  structures, or a perturbation of the complex array element excitation weights~\cite{DingFusco1}.

In many DM systems, the orthogonal vector approach introduced in~\cite{DingFusco2014,DingFusco2} is employed. When we use this paradigm, a DM implementation is a modification of array excitation weights corresponding to a $\pm \pi/2$ rotation; this is similar to the artificial noise approach utilized by the information theory community \cite{NegiGoel1,NegiGoel2}. These induced rotations are balanced, such that the cumulative transmission arrives on-phase. This results in a necessary trade of power placed on target by the transmitting platform, e.g. half of the elements experiencing a rotation leads to a roughly 6 dB loss, in exchange for the increased security off-target. However, this presents a problem that becomes apparent in the limit of few-element arrays. This comes about from the limited set of unique combinations of conventionally-transmitting and randomly-rotated nodes. For a capable enough eavesdropper, the distorted symbols may be recoverable as was demonstrated in \cite{Hafez2016}.

A simple extension of the DM concept is presented in \cite{Fan2018} in which the random phase rotations induced by the array excitation weights are chosen to be real-valued and small. This technique induces the desired security aspects contained in DM, while creating a larger set of unique combinations; however, this approach induced some variability in the received power at the target location. Other approaches utilize genetic algorithms~\cite{DalyBernhard2} or particle-swarming~\cite{DingFusco3} methods to perform the optimization task of selecting the array weights. The use of these optimization approaches can be prohibitively expensive in terms of computational cost, especially when the transmitting system is expected to perturb array weights on a per-symbol basis. In-between these two families of approaches is where our work is placed; providing a computationally-efficient means to randomly perturb the individual array weights such that security off-target is increased yet providing enough constraint so that the energy delivered on-target is constant, or for amplitude modulation schemes, matching the desired amplitude.

\IEEEpubidadjcol
This paper is organized as follows. In Section~\ref{section2}, we provide the preliminary mathematical groundwork necessary for the continued discussion and present the operational scenario that guides the rest of our discussion. We then demonstrate the functionality of a conventional DM framework. Section~\ref{section3} discusses the real-valued phase approach to DM, wherein we also discuss modifications to further extend this capability and provide broad operational relevance. Results that provide direct comparison between conventional and real-valued phase DM is discussed in Section~\ref{section4} with final statements following in \ref{section5}.
\IEEEpubidadjcol
\end{section}

\section{Background}\label{section2}
Herein we present the salient mathematical features of DM to aid in denoting the differing, noteworthy, behaviors between conventional and real-valued phase approaches to DM. This discussion is concluded with the operational scenario that will be used as the exemplar for all future discussions within this publication that discuss performance benefits and differences between conventional and real-valued phase DM.

\subsection{Mathematical Description}
We assume isotropically-radiating antenna elements with uniform array spacing of one half wavelength. The far-field pattern, $\mathbf{E}(\theta)$, is determined as a sum of the individual element contributions with the respective element excitations $\alpha_n$,
\begin{equation}\label{ff}
  \mathbf{E}(\theta)=\sum_{n=1}^N(\alpha_n e^{i\pi(n-\frac{N+1}{2})cos(\theta)}).
\end{equation}
The implementation of DM is then achieved through replacing the individual element excitations with the product of the modulated data stream and complex gain of unity \todo (context, options, simplicity??) magnitude and non-zero phase in the IQ-space, such that: $\alpha_n \equiv d_m G^*_{n}$ and $G_{n} = e^{j\phi_n}$.
\par
Moving from a single epoch to multiple epoch processing of DM sequences introduces the requirement of constraining orthogonality across the gain \textit{differences} over the array. Consider two epochs, wherein the gain coefficients can be represented via vectors $\vec{G}_u$ and $\vec{G}_v$ respectively. The gain difference is the simply, $\Delta \vec{G}=\vec{G}_u-\vec{G}_v$. Then, discretizing over the spatial direction in Eq.~\ref{ff} we can denote the discrete channel matrix as,
\begin{equation}\label{ch}
 \mathbf{H}\equiv
 \begin{bmatrix}
  e^{i(1-\frac{N+1}{2})\pi cos(\theta_1)} & \hdots & e^{i(N-\frac{N+1}{2})\pi cos(\theta_1)} \\
  e^{i(1-\frac{N+1}{2})\pi cos(\theta_2)} & \hdots & e^{i(N-\frac{N+1}{2})\pi cos(\theta_2)} \\
  \vdots & \ddots & \vdots \\
  e^{i(1-\frac{N+1}{2})\pi cos(\theta_K)} & \hdots & e^{i(N-\frac{N+1}{2})\pi cos(\theta_K)}
 \end{bmatrix}.
\end{equation}
We then can note the orthogonality constraint, as presented in \cite{DingFusco2}, as
\begin{equation}
 \mathbf{H}(\theta_k = \theta_{target})^*\Delta\vec{G} = 0~.
\end{equation}

\par
We note the primary difference between conventional and real-valued phase DM. Conventional DM systems, such as those defined in \cite{DingFusco1,DalyBernhard2} constrain the phase of the complex gain to three values $\phi_n = [-\pi/2, 0, \pi/2]$. Whereas in the real-valued phase DM framework, we denote $\phi_n$ as a random variable drawn from a beta distribution in the interval $\phi_n \in [-\pi/2, \pi/2]$, that is $\phi_n \sim \frac{\pi}{2}[\Be(\alpha=\beta)-.5]$. We present the gain determination algorithm in Algorithm~\ref{alg:alg1}.\par
\begin{figure}[h!]
 \centering
 \includegraphics[width=.75\columnwidth]{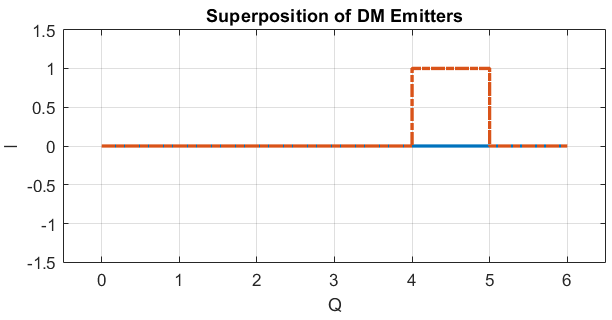}
 \caption{Illustrative example of a vector path obtained through an array excitation approach consistent with conventional DM}
 \label{superpos}
\end{figure}

The implications of our modifications to the DM framework are more readily assessed in the framework of vector synthesis presented in \cite{DingFusco2,DingFusco2014}. Wherein we examine the received symbol as a superposition of vector contributions from the individual emitters in IQ space; see Fig.~\ref{superpos} where we present a simple case of conventional DM with an eight element ULA.
\begin{figure}[h!]
 \centering
 \includegraphics[width=.75\columnwidth]{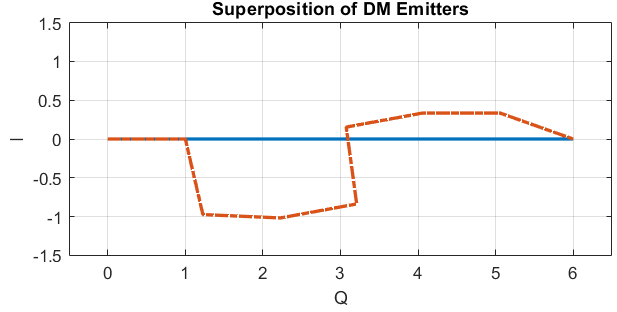}
 \caption{Vector path in IQ-space obtained through the real-valued phase DM approach.}
 \label{superpos2}
\end{figure}\\
Here we observe that two of the eight emitters performed a $\pm\pi/2$ rotation resulting in an approximate 2.5 dB loss. It is an easy exercise to note the difficulty in finding unique paths through IQ space induced by the cumulation of orthogonal array contributions when the array has relatively few elements. Within this problem space, we can easily denote the benefit of the real-valued phase framework for DM. In Fig.~\ref{superpos2} we demonstrate the effect that the real-valued phase rotations has on the vector path in IQ-space. It is important to note the behavior of the final two vector contributions, where an optimized close-out procedure is performed to reach the appropriate symbol phase and amplitude. This result is achieved through assessing the geometry of the vector path and constructing and appropriate equilateral triangle with the final element contributions.

\subsection{Permutation}
\begin{figure}[h!]
 \centering
 \includegraphics[width=.9\columnwidth]{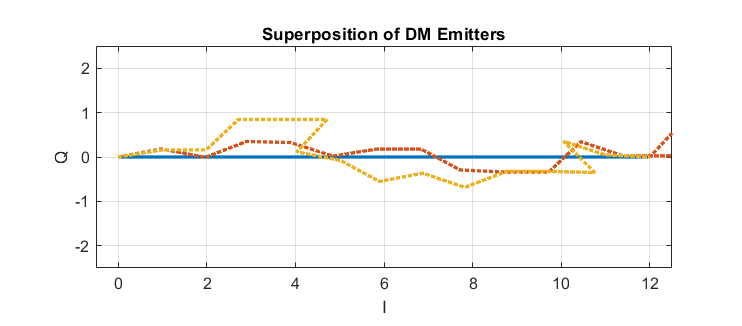}
 \caption{Vector path of permuted gain sequences in the real-valued phase approach to DM.}
 \label{permute}
\end{figure}

As we briefly discussed previously, the motivation for including real-valued phase rotations in the gain determination for DM beamforming was to increase the configuration diversity. This \todo (decreases???) increases the periodicity of repetition in DM transmissions which decreases the likelihood of an eavesdropper accumulating sufficient information to decode the transmission and to effectively \todo(repetition, break?) decode the DM. In the limit of large, dense, phased arrays it may become too cumbersome to repeatedly compute the individual phase rotations at the symbol rate. In this situation it is important to note that permutations and reflections about the symbol axis in IQ-space can be performed for low-cost determination of additional sequences of gain coefficients across the array. As an example, consider the case presented in Fig.~\ref{permute} which shows a permuted case of vector paths for DM utilizing the real-valued phase framework in our exemplar presented above.

\subsection{Exemplar}
For ease of analysis we consider a linear phased line array (ULA) with operational frequency of 1 GHz. The phased array consists of $N=16$ elements separated by the conventional $\lambda/2$ spacing. Each element is transmitting with identical, unity, gain and is steering the main lobe response at $0^\circ$ from the array normal. We define eavesdroppers as receiving platforms located at angular positions outside of the main lobe response. The scenario embodies a dynamic scene with stationary transmitter and receiver platform(s), i.e. symbol rate differences are implied in the transmit scheme. Within each epoch the DM framework must match a symbol corresponding to $0^\circ$ phase and amplitude $75\%$ of total array power, that is an approximate 2.5 dB loss in intended recipient direction when compared to conventional beamforming techniques.

\subsection{Conventional DM}
We can apply conventional DM to the exemplar presented above and examine the performance of the transmitting platform over a time-varying scenario. This allows us to examine the variability that the DM paradigm is capable of introducing in the radiation placed off-target. This is most readily observable in the variance and average of the radiation pattern over several epochs, signified by the generation of a new set of phase rotations.
\begin{figure*}[t!]
 \centering
 \includegraphics[width= 7.5cm,height=5cm,keepaspectratio]{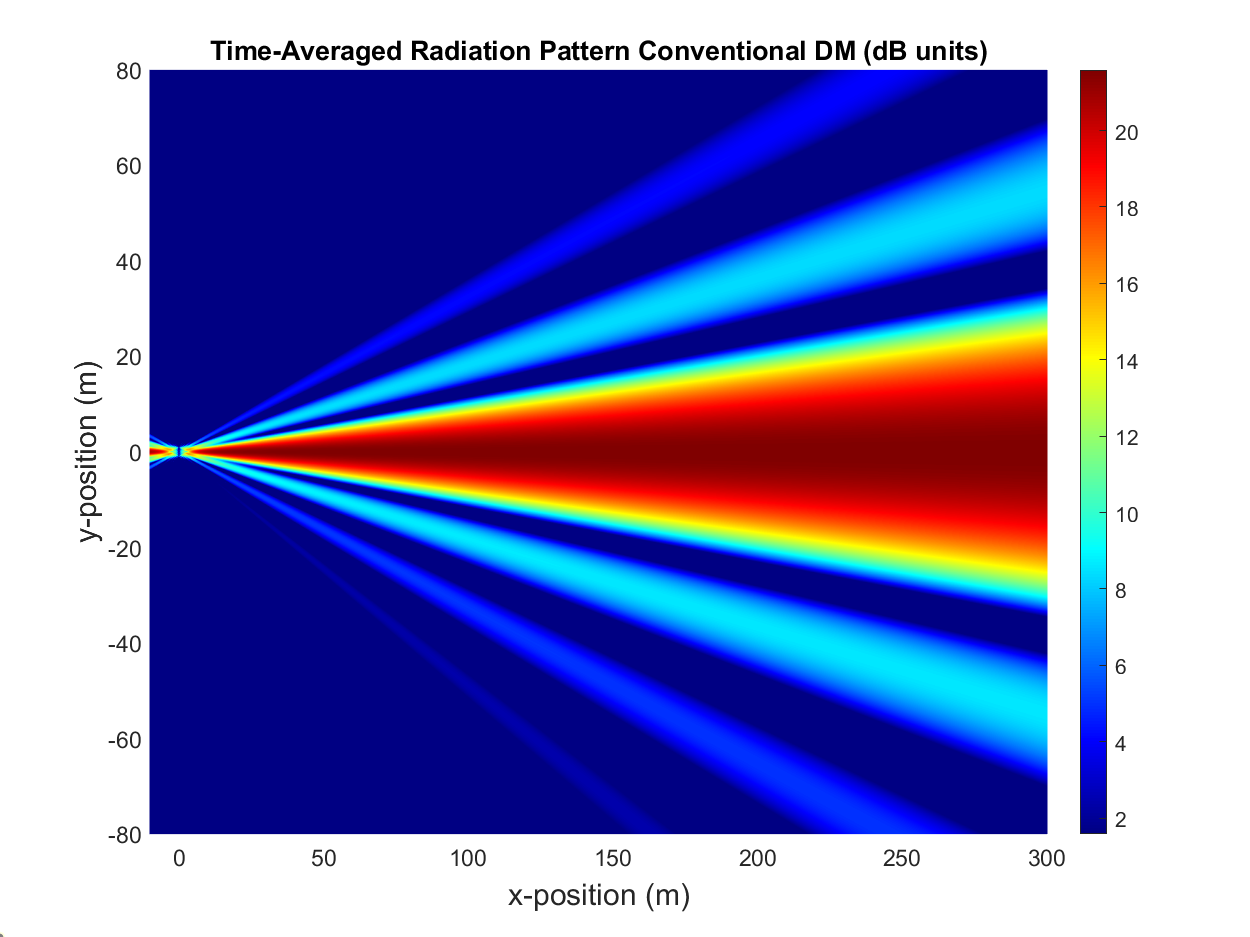}
 \includegraphics[width= 7.5cm,height=5cm,keepaspectratio]{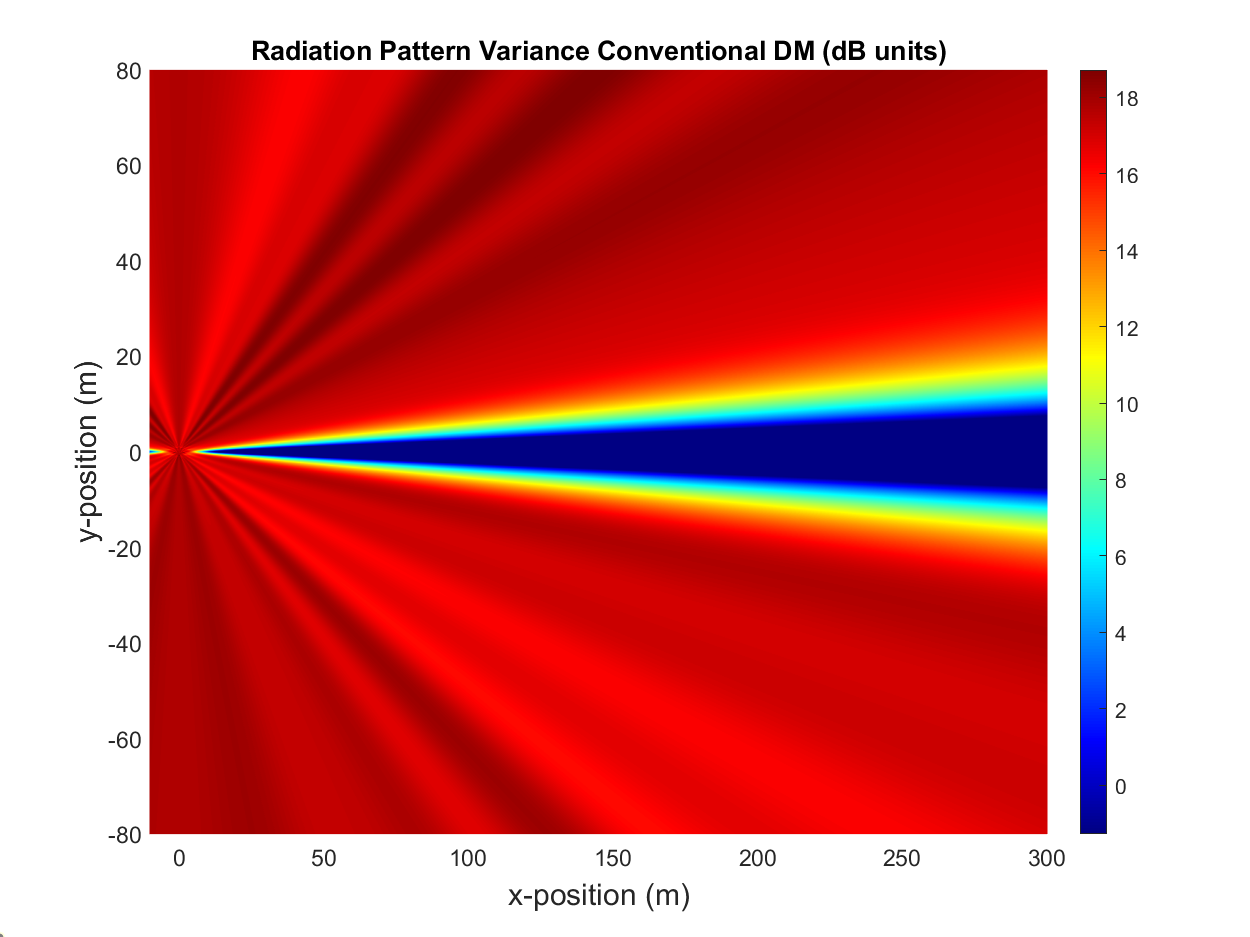}
 \caption{Time average (left) and variance (right) of radiation pattern for a 16 element ULA performing conventional DM over 200 epochs.}
 \label{convDMRad}
\end{figure*}
We present this result for conventional DM in Fig.~\ref{convDMRad} over an operational period comprising 200 independent transmission epochs. The results demonstrate the ability of the conventional DM framework to disrupt eavesdropper capability of interception by causing the received energy to be unreliable over time in directions that do not correspond to the intended direction. All the while, the variance vanishes as we approach the intended transmission direction which is a consequence of the orthogonality constraint and implying no, or limited, impact for the intentional link as well.

\begin{figure*}[b!]
\centering
\includegraphics[width= 15cm,height=5cm,keepaspectratio]{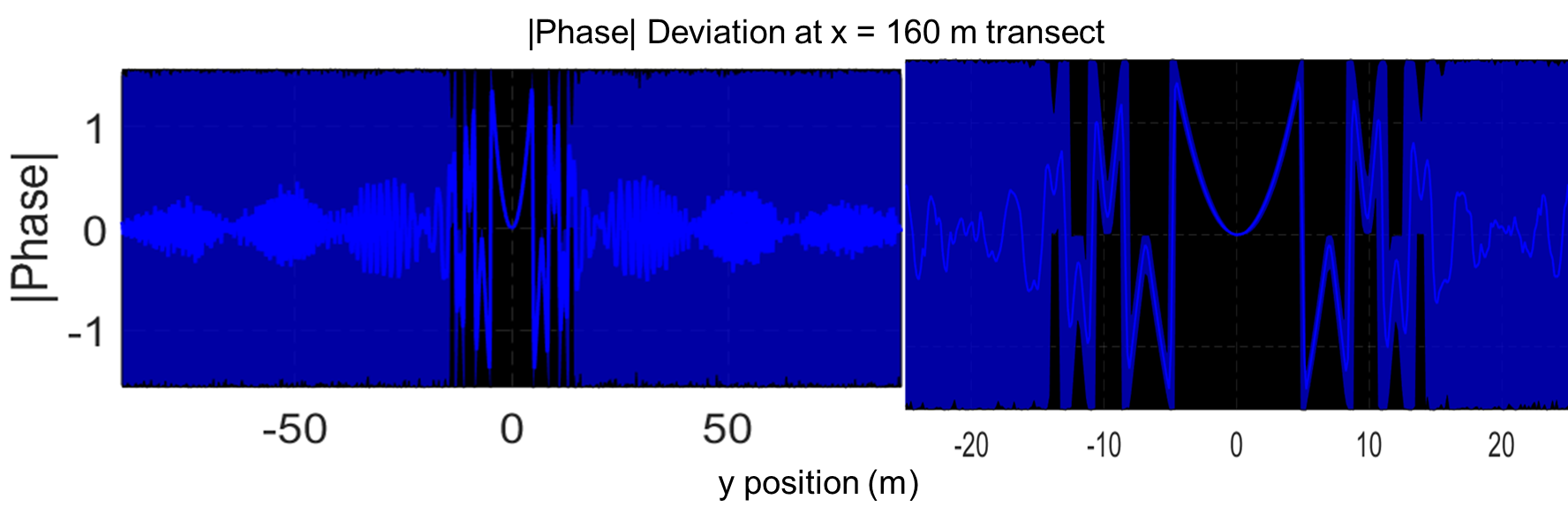}
\caption{Phase deviation in vertical transect through target location showing the mean and min/max deviation from the average.}
\label{phaseConv}
\end{figure*}\par
The results contained in Fig.~\ref{convDMRad} are bolstered by the ability of conventional DM to provide epoch-to-epoch scrambling of phase for the symbol location in the constellation. We provide this for conventional DM in Fig.~\ref{phaseConv} where we observe a full $\pi$ of symbol phase variability at solid angles of approximately $11^\circ$.

\begin{section}{Performance of Real-Valued DM}\label{section3}
With the results of the conventional DM approach in mind, we now turn our attention to the results generated with the real-valued phase DM framework. We restrict our analysis to the same exemplar that we used for our discussion of conventional DM. The beta distribution used for the real-valued DM application was parameterized as $\alpha=\beta=3$. We see directly in Fig.~\ref{revalDMRad} that the real-valued phase DM framework is capable of providing the same phenomena of variable energy delivered in the side lobes. This comes at the cost of degree of variability for the delivered energy when compared to the result for conventional DM.

\begin{figure*}[t!]
 \centering
 \includegraphics[width= 8cm,height=8cm,keepaspectratio]{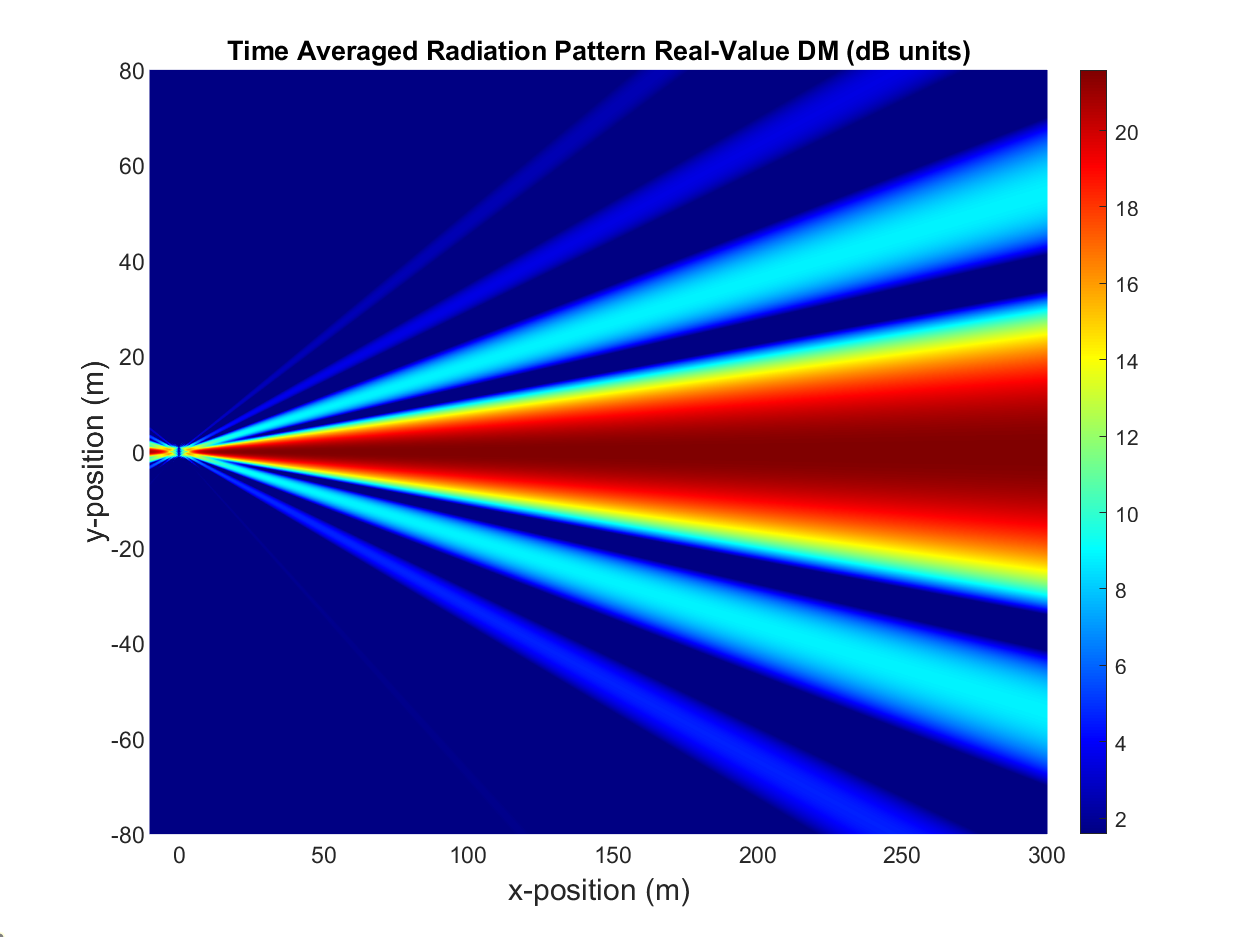}
 \includegraphics[width= 8cm,height=8cm,keepaspectratio]{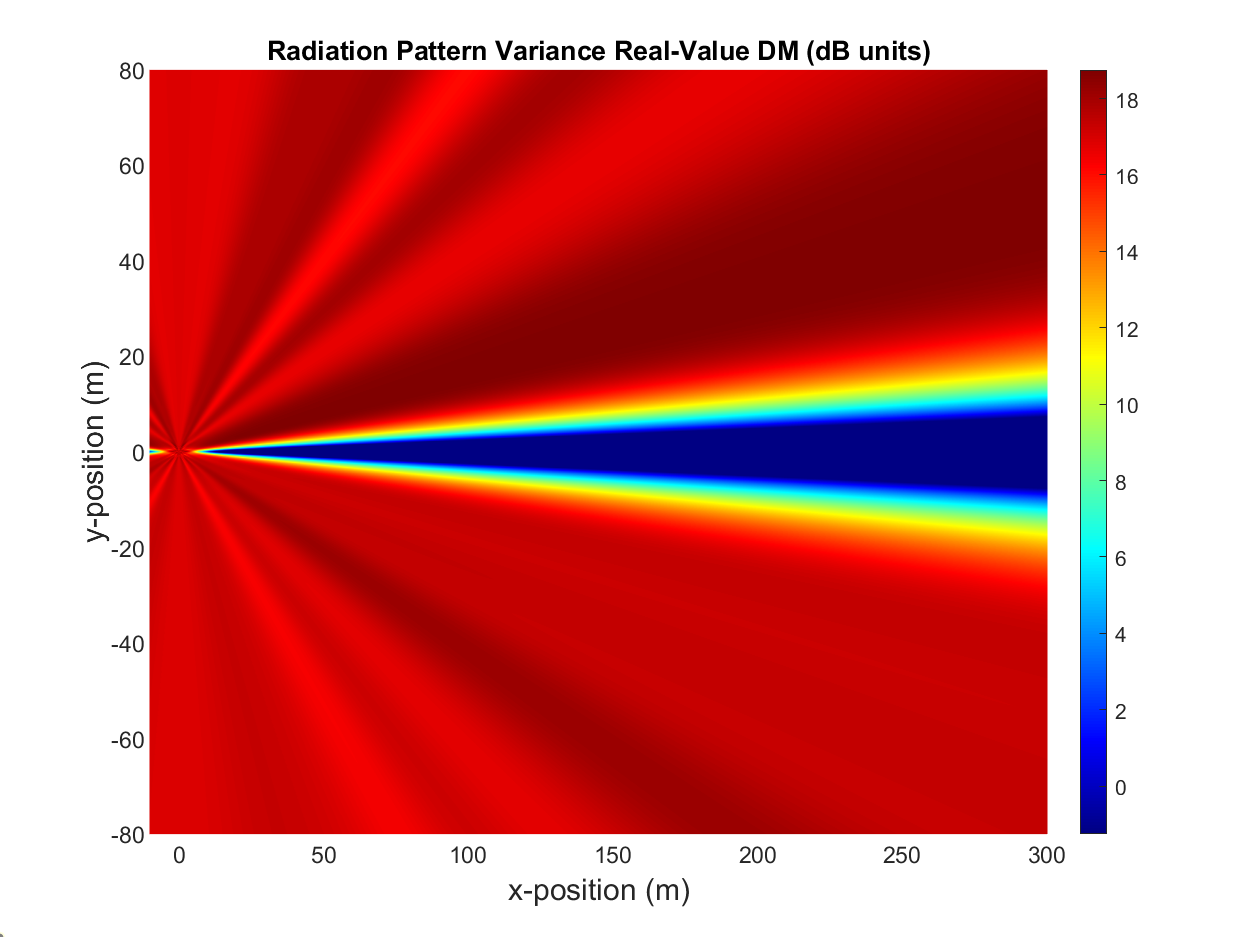}
 \caption{Time average (left) and variance (right) of radiation pattern for a 16 element ULA performing real-valued DM over 200 epochs.}
 \label{revalDMRad}
\end{figure*}


\subsection{Convergence}
As described in algorithm \ref{alg:alg1}, we could rely upon repeated pulls from the random number generator to arrive at an $N-c$ sized set of phases whose element weights' sum comes within $c$ of the desired amplitude at zero phase.
To guarantee that the algorithm converges to the desired sum without repeated pulls from the random number generator, we can inspect the cumulative sum of the weights. When the euclidean distance between the cumulative sum and symbol location drops below the number of remaining elements to inspect, i.e. less than $c$. When this occurs the remaining weights are replaced by unit vectors pointing in IQ-space directly from the cumulative sum to the destination. This can be thought of as allowing the random walk vector to do as it will until it is sufficiently far off the path to the destination that randomness is set temporarily aside to herd the vector path towards our goal. When the cumulative sum of the vector path is within $c$ of the desired amplitude at zero phase with just $c$ weights remaining, the final $c$ weights are calculated instead of randomly chosen. The $c$ complex vectors form an equilateral triangle with the vector from the tip of the vector path towards the target amplitude serving as the base. We term the calculation of these $c$ weights as close out in the algorithm description, as seen in Fig.~\ref{algoflow}
\begin{algorithm}[H]
\caption{Complex Gain Determination with Unit Amplitude}\label{alg:alg1}
\begin{algorithmic}
\STATE
\STATE {\textsc{Assign Phases:}} $\phi_n$
\STATE \hspace{0.5cm}$ \phi_n \sim \frac{\pi}{2}[\Be(\alpha=\beta>1)-.5]$
\STATE {\textsc{Assign Values for Closeout}}
\STATE \hspace{0.5cm}$ \vec{\mathcal{H}}_{co} = \sum_n^{N-c}\vec{G}_n $
\STATE \hspace{0.5cm}$ \vec{A}_{offset} = \vec{A}-\vec{\mathcal{H}}_{co}$
\STATE \hspace{0.5cm}$ \phi_{offset} = -tan^{-1}(\frac{\mathfrak{Im}(\vec{\mathcal{H}}_{co})}{\mathfrak{Re}(\vec{\mathcal{H}}_{co})})$
\STATE {\textsc{IF}:}$ ||\vec{A}_{offset}||_2 < 2 $
\STATE \hspace{0.5cm} $\vec{G}_N = \frac{1}{2}\vec{A}_{offset} +\hat{A}_{offset}e^{i\pi/2}\sqrt{1-(\frac{1}{2}|\vec{A}_{offset}|)^2}$
\STATE \hspace{0.5cm} $\vec{G}_{N-1} = \vec{A}_{offset} - \vec{G}_N$
\STATE {\textsc{ELSE IF}:}$ ||\vec{A}_{offset}||_2 = 2 $
\STATE \hspace{0.5cm} $\vec{G}_N=\vec{G}_{N-1}=e^{i\phi_{offset}}$
\STATE {\textsc{ELSE}:}
\STATE \hspace{0.5cm} {\textsc{REDRAW}}
\STATE \textbf{return}  $\{\vec{G}_n:n=1,..,N\}$
\end{algorithmic}
\end{algorithm}
\begin{figure}[h!]
 \centering
 \includegraphics[width=.75\columnwidth]{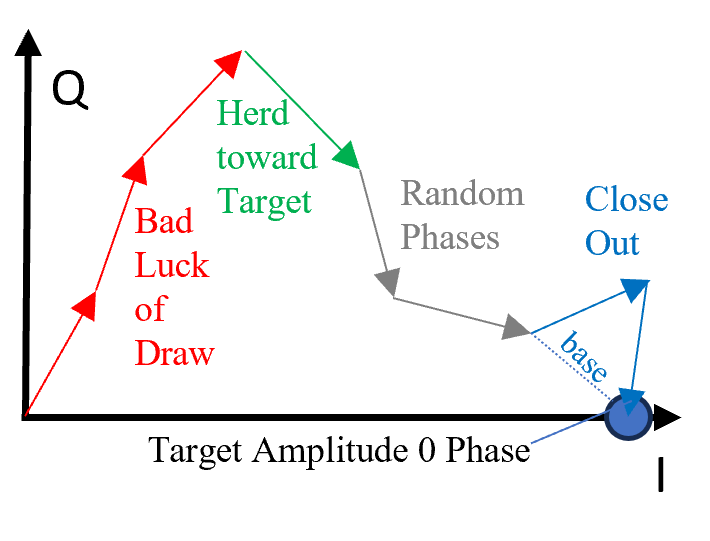}
 \caption{Illustrative example of a vector path with a random weight replaced with one calculated to head directly toward the target (green) and with the final two weights forming an equilateral triangle (blue)}
 \label{algoflow}
\end{figure}
Ultimately, the choice of PDF permits more than a scaled and shifted symmetric Beta distribution; trades can be made to lower the computational complexity of the random number generator with other PDFs. The algorithm is generally agnostic to this choice, so it can be taken as a design consideration. Of course, there are limits in the ability to guarantee convergence with the selection of the goal amplitude. A 16 element array can only provide maximal directive field strength gain of 16x. Since DM requires some backoff from maximum gain to provide space to dither the weights, the algorithm requires some backoff as well. In the case of 16 elements, the target amplitude probably shouldn't be 15.99; although it is often easily greater than approximately half, as is seen in many traditional DM implementations.

\subsection{Random Phase Distribution Functions}
In the default configuration, Algorithm 1 draws from a scaled and shifted version of a symmetric Beta distribution. We prefer the symmetric Beta distributions to generate the random phases for several reasons. Of course, as is built into its name, is the symmetry about a peak such that the expectation of the sum of the phases can be posed to reliably provide zero after a simple shift of the distribution. The impact of scaling makes it such that its range of support is confined to the finite interval bordered by $\pm \pi/2$. This is in comparison to a uniform distribution which is similarly shifted and scaled, wherein the bulk of its density is focused on values closer to zero but exists on an open interval. When considering the aforementioned guarantee of converging on one pull of a set of phases, the emphasis on small phases makes it less likely that the cumulative sum needs to be modified to herd the phase toward the target amplitude with zero phase.

\subsection{Amplitude Modulation with Constant Envelope Drivers}
Regardless of the means, DM provides a capability to meet quantized amplitude values through spatial incoherence in directions not representing intended transmit locations. This feature is necessary for arrays driven by amplifiers that require unity peak to average power ratios. This can be used to meet standard QAM schemes for arbitrary symbol constellations. This capability is shown in Fig.~\ref{AmpMod} with a 4-ary amplitude modulation scheme used as the example. For each amplitude level, the target amplitude is adjusted so that the summation of weights in the intended direction receives the appropriate amplitude. This is not a unique feature of real-valued phase DM as the number of $\pm \pi/2$ weights in traditional DM could also be varied to adjust the aggregate amplitude.
\begin{figure}[h!]
 \centering
 \includegraphics[width=9cm, height=9cm,keepaspectratio]{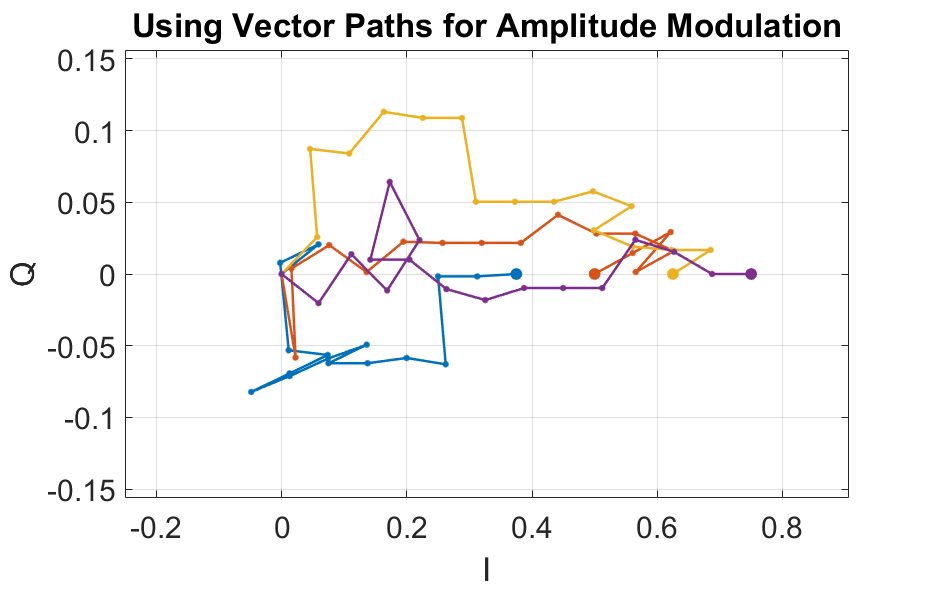}
 \caption{Demonstration of achieving variable amplitude modulation schemes through the use of vector paths induced by real valued phased DM.}
 \label{AmpMod}
\end{figure}

\subsection{Relaxing Unity Gain Constraint}
Maintaining the constraint of constant, unity, magnitude for the gain coefficients of the individual emitters allows us to be flexible to a wider variety of platforms. However, this constraint can be argued to be unnecessary for a large range of scenarios; characterized by the capability of each element, or individual transmitter, to emit with variable gain magnitude within the interval $0 < G_n \leq 1$. This framework has equivalent metrics in terms of unique paths through IQ-space, represented in Fig.~\ref{superpos3}. The algorithm for determining the complex gain coefficients is described in Algorithm~\ref{alg:alg2}.
\begin{figure}[h!]
 \centering
 \includegraphics[width=9cm, height=4cm,keepaspectratio]{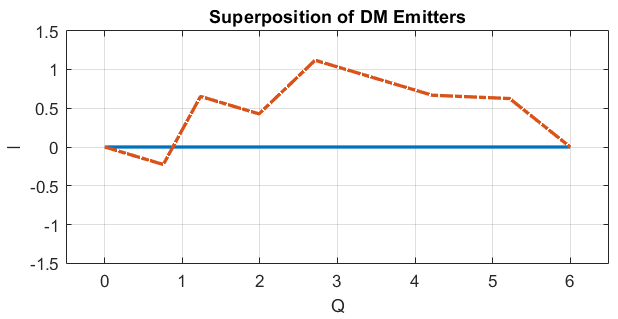}
 \caption{Vector path in IQ-space obtained through the real-valued phase DM approach with each emitter capable of transmitting with variable amplitude gain.}
 \label{superpos3}
\end{figure}
\begin{algorithm}[H]
\caption{Complex Gain Determination with Variable Amplitude}\label{alg:alg2}
\begin{algorithmic}
\STATE
\STATE {\textsc{Assign Phases:}} $\phi_n$
\STATE \hspace{0.5cm}$ \phi_n \sim \frac{\pi}{2}[\Be(\alpha=\beta>1)-.5]$
\STATE {\textsc{Assign Values for Closeout}}
\STATE \hspace{0.5cm}$ \vec{\mathcal{H}}_{co} = \sum_n^{N-c}\vec{G}_n $
\STATE \hspace{0.5cm}$ \vec{A}_{offset} = \vec{A}-\vec{\mathcal{H}}_{co}$
\STATE \hspace{0.5cm}$ \phi_{offset} = -tan^{-1}(\frac{\mathfrak{Im}(\vec{\mathcal{H}}_{co})}{\mathfrak{Re}(\vec{\mathcal{H}}_{co})})$
\STATE \hspace{0.5cm}$\vec{G}_N = \frac{||\vec{A}_{offset}||_2}{c}\big[cos(\phi_{offset})+isin(\phi_{offset})\big]$
\STATE \textbf{return}  $\{\vec{G}_n:n=1,..,N\}$
\end{algorithmic}
\end{algorithm}
\end{section}

\begin{section}{Results}\label{section4}
When we discuss the benefit of real-valued phase DM the argument is centralized around two concepts, the uniqueness of the sequence of complex gain coefficients and the benefit of rich permutation sets of possible gain coefficients towards increased secrecy capacity. Having discussed the algorithmic differences in the two approaches we can then examine how our proposed changes impact our performance by these quantifiable measures.
\subsection{Uniqueness of Gain Coefficients}
\begin{figure}[h!]
 \centering
 \includegraphics[width=1\columnwidth]{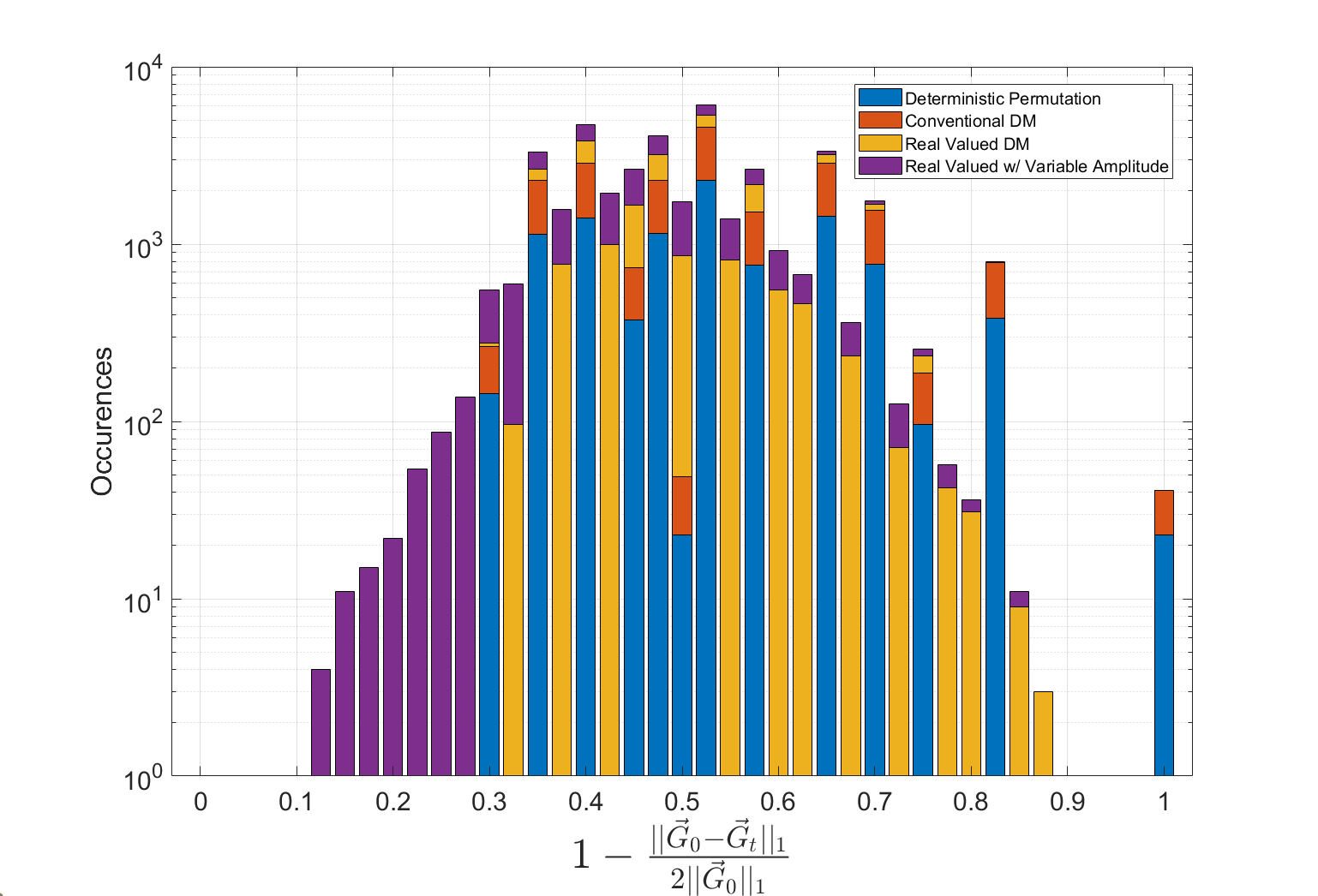}
  \caption{Uniqueness over time of orthogonal noise vector for the various DM frameworks considered herein. Result shows the repetitions in conventional DM weights from the initial epoch considered as reference through numerous counts of unit value.}
 \label{unique}
\end{figure}
\begin{figure*}[t!]
  \centering
 \includegraphics[width= 8cm,height=4cm,keepaspectratio]{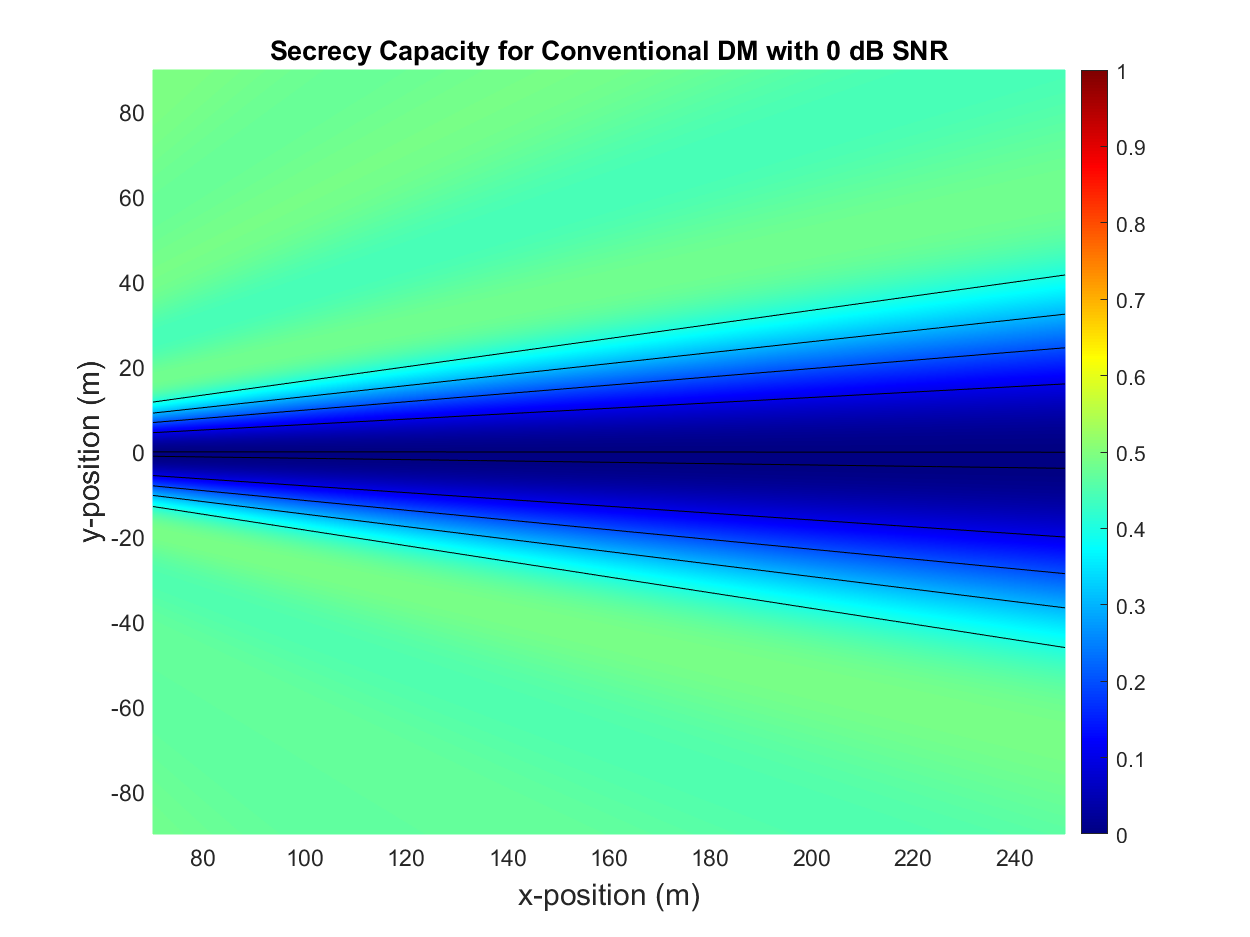}
 \includegraphics[width= 8cm,height=4cm,keepaspectratio]{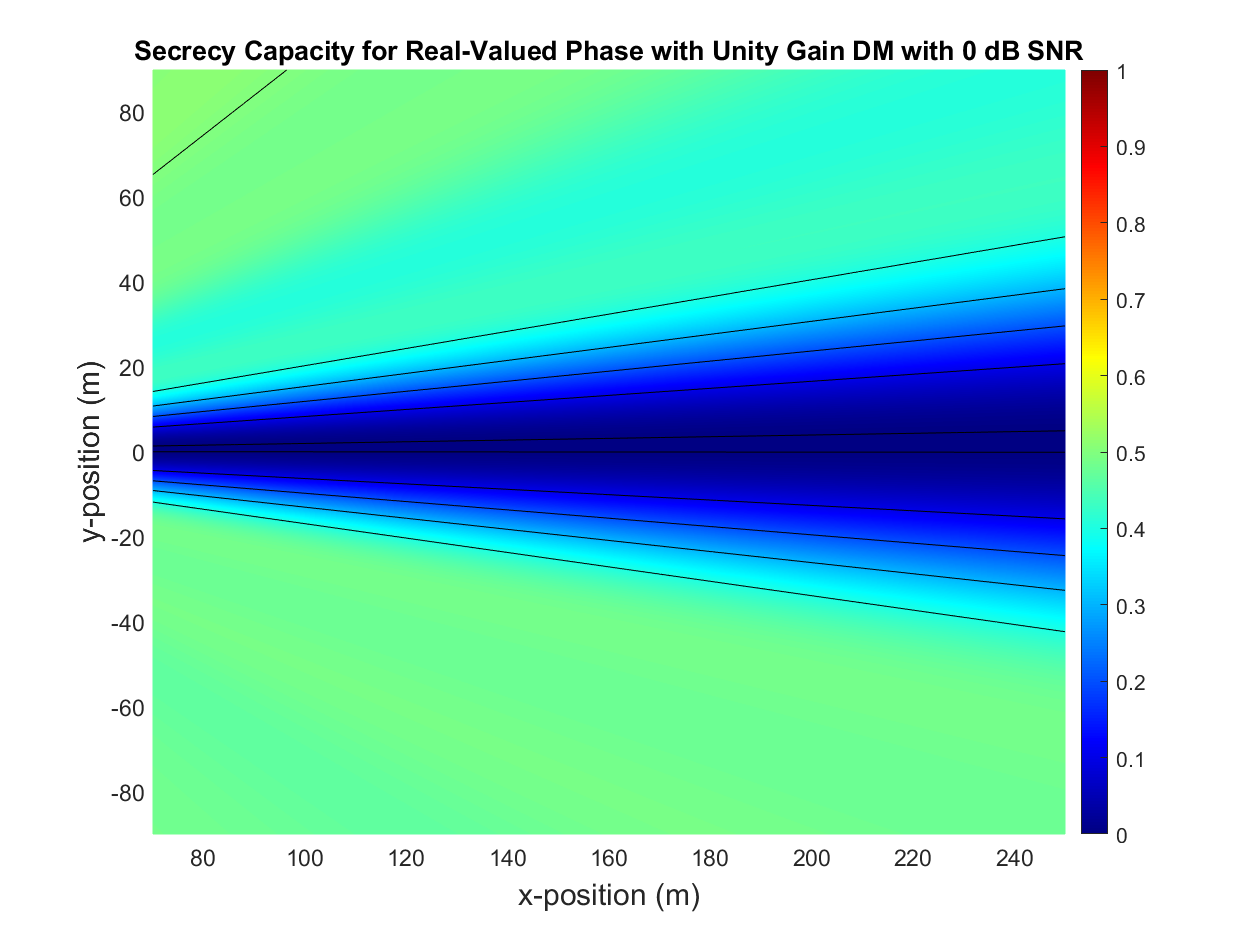}
 \includegraphics[width= 8cm,height=4cm,keepaspectratio]{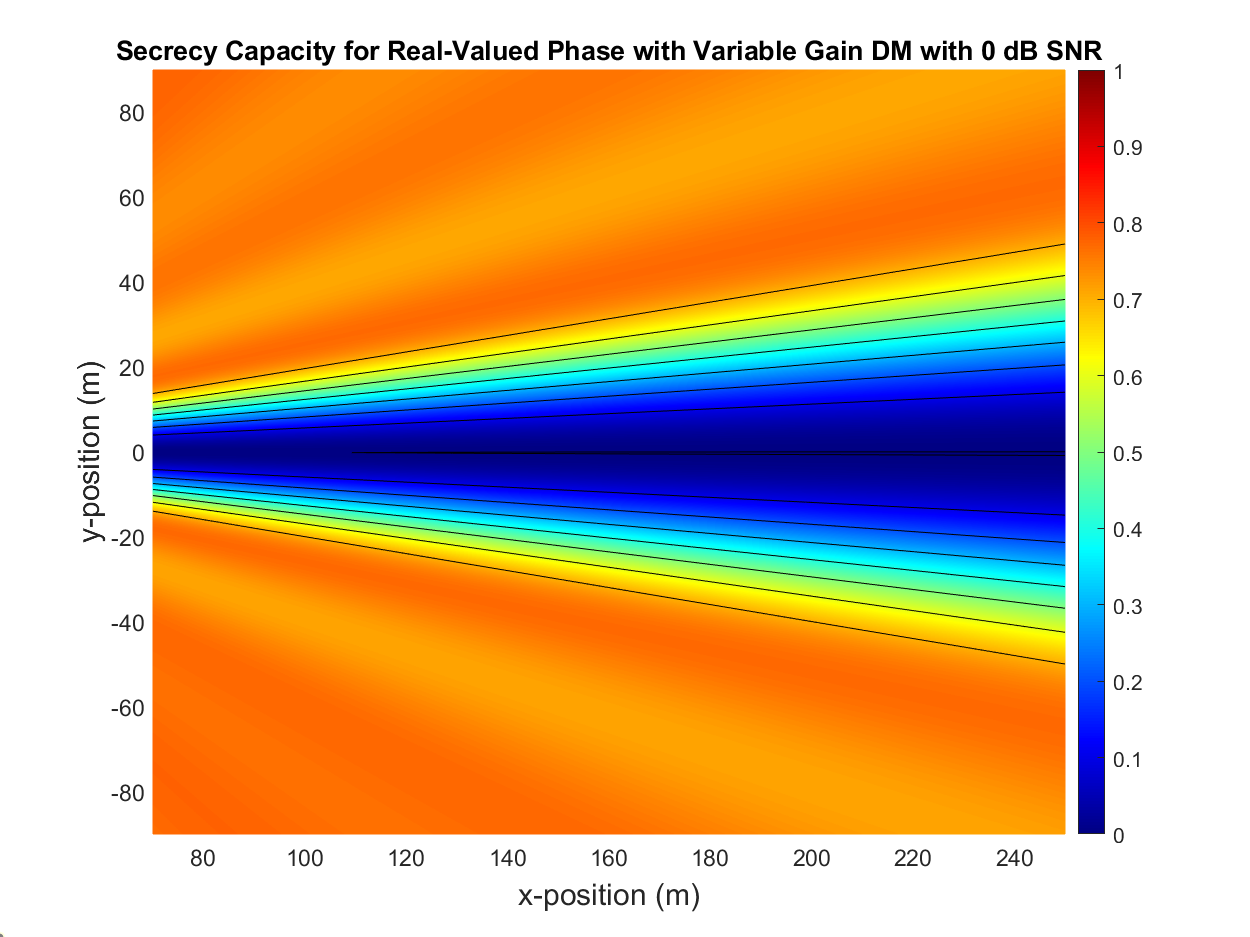}
 \caption{Secrecy spectral efficiency for conventional (left), real-valued phase with unit amplitude (middle), and real-valued phase with variable amplitude (right) DM for an 8-element ULA with 4 randomized phase rotations in the absence of noise and over 500 epochs.}
 \label{seccap}
\end{figure*}
The modification of DM to include real-valued phase rotations was done to increase the availability of unique vector paths through IQ-space. Utilizing the orthogonal noise injection framework introduced by \cite{DingFusco2014} we denote the possibility of a false positive for orthogonality checks when the sequence of gain coefficients is repeated from a prior epoch. This issue is amplified with the use of arrays which consist of only a few elements. This is due to the finite length of the set of unique permutations for the gain coefficients of an array with $N$ elements and utilizing $k/2$ rotations of $\pi/2$ and $-\pi/2$ each. The size of this set scales via permutations with coupled choices, i.e. as $\frac{N!}{(N-k)!(k/2)!(k/2)!}$. By considering the possibility of real-valued phase rotations for the complex gain coefficients the length of our set of unique gain coefficients is reinforced by the set of quantizable phase rotations that are permissible by the bit-value of the appropriate precision. To analyze this, we consider a scenario of an 8-element phased array. Of the 8 elements in the original array, we permit phase rotations in half; these phase rotations are equally split among $\pm\pi/2$ to guarantee no symbol distortion in the intended direction. The set of permutations for this configuration is then $\frac{8!}{4!2!2!}=420$ which corresponds to the periodicity of a particular DM configuration, or more succinctly how many unique DM transmissions can occur with the given array. We can observe this behavior by looking at the Manhattan distance for real-valued phases, of a sequence of gain coefficients with following permutations over the processing period. This is shown for a simulated period of 10000 epochs. This allows for adequate coverage of the period for conventional DM; we present this result for conventional DM, real-valued phase DM, and real-valued phase DM with variable amplitude in Fig.~\ref{unique}. We see in this result that, over the period considered, conventional DM has two repetitions of the full initial sequence of gain coefficients which can be seen in Fig.~\ref{unique}. To be thorough and lend rigor to our analysis we also present the case of permuted, conventional DM; wherein the complex gain coefficients are methodically selected from the full set of permutations as we discussed previously.

\subsection{Secrecy Capacity}

Metrics that are useful for assessing the performance of DM systems are discussed in \cite{DingFusco4}, among them is the notion of secrecy capacity \cite{Shannon,Wyner,BarrosRodr,CsisKorn} which is widely used in the information theory community as a measure of the maximum amount of information that a receiver can reliably recover from a transmission that remain unrecoverable by an eavesdropper. This is denoted as the difference in link capacity between the legitimate listener and eavesdropper, $C_s = (C_r -C_e)^+$. As the eavesdroppers capacity decreases, unintended listeners lose the ability to decode transmissions intended for the legitimate user. Herein we use the term secrecy capacity for what is truly secrecy spectral efficiency, the ratio of secrecy capacity to bandwidth, with units of bits/second/Hz. Starting from the formulation of the radiation pattern described in Eqn.~\ref{ff} we can denote the received signal at angular location $\theta$ for a discrete time index $m$ as,
\begin{equation}
 y(\theta,m) = h_m(\theta)x(m)+\eta = \mathbf{E}(\theta,m) + \eta~.
\end{equation}
Where $\eta$ is the complex, zero-mean Gaussian noise present in the channel and we have introduced the complex fading coefficients $h_m(\theta)$ and modulated data stream $x(m)$. The instantaneous received SNR at an angular location $\theta_K$ for a receiver operating with noise power $N(\theta)$ is then,
\begin{equation}
 \gamma_m(\theta_K) = \frac{||\mathbf{E}(\theta_K,m)||^2}{N(\theta)}
\end{equation}
Averaging the SNR over a processing window allows us to introduce manipulations on the time-wise variance of the radiation pattern. Denoting the expectation and variance operators as $\mathbb{E}\big[.\big]$ and $\mathbb{V}\big[.\big]$ respectively and the time-averaged noise power as $\bar{N}(\theta)$, we obtain simply
\begin{equation}
 \mathbb{E}\big[\gamma_m(\theta)\big] = \frac{\mathbb{E}\big[||\mathbf{E}(\theta_K,m)||^2\big]}{\bar{N}(\theta)}~.
\end{equation}
Noting the definition of the variance of a distribution,
\begin{equation}
 \mathbb{V}[x] \equiv \mathbb{E}[x^2]-\mathbb{E}[x]^2.
\end{equation}

Which allows us to rephrase the time-averaged SNR in terms of the qualitative results demonstrated in Figs.~\ref{convDMRad} and~\ref{revalDMRad}
\begin{equation}
 \mathbb{E}\big[\gamma_m(\theta)\big]=\bigg\{\mathbb{V}\big[\mathbf{E}(\theta)\big]+\mathbb{E}\big[\mathbf{E}(\theta)\big]^2\bigg\}/\bar{N}(\theta)~.
\end{equation}

Then, as is discussed in \cite{BarrosRodr} and \cite{Hafez2016}, we can denote the spatial distribution of the achievable secrecy rate through utilizing the time-averaged SNR,
\begin{equation}
 R(\theta) = {\rm log}_2\bigg(1+\mathbb{E}\big[\gamma_m(\theta)\big]\bigg)~.
\end{equation}
With the total secrecy capacity of the transmission denoted as the diferrence in secrecy rate between the intended direction $\theta_r$ and the arbitrary location of a eavesdropper $\theta_e$, i.e.
\begin{align}
\nonumber C_s &= R(\theta_{r})-R(\theta_{e})\\
&= {\rm log}_2\bigg\{\bigg(\frac{\bar{N}(\theta_e)}{\bar{N}(\theta_r)}\bigg)\frac{\bar{N}(\theta_r)+\mathbb{V}\big[\mathbf{E}(\theta_{r})\big]+\mathbb{E}\big[\mathbf{E}(\theta_{r})\big]^2}{\bar{N}(\theta_e)+\mathbb{V}\big[\mathbf{E}(\theta_{e})\big]+\mathbb{E}\big[\mathbf{E}(\theta_{e})\big]^2}\bigg\}~.                                                                                                                                                      \end{align}
The comparison of the secrecy capacity of the two DM frameworks discussed herein are presented in Fig.~\ref{seccap} in a scenario with 0 dB SNR over a 500 epoch processing window. We can see that the real-valued phase approach to DM with variable gain drastically improves the secrecy capacity of the transmission. This secrecy capacity estimate in Fig.~\ref{seccap} is only based by changes in SNR as a matter of beamforming; note the similarity with Fig.~\ref{convDMRad}. Secrecy capacity is further improved when $ \bar{N}(\theta_r) $ increases due to the noise-like distortion of directional modulation.

\end{section}

\begin{section}{Conclusion}\label{section5}
We have presented the underlying framework for performing directional modulation (DM) on phased arrays by considering the process as that of selecting the complex weight coefficients leading to a constrained, randomized, walk of vector path accumulations through IQ-space. We showed that in conventional DM frameworks, where the modifying weight coefficients are limited to $\pm \pi/2$, selecting unique permutations of the sequence of gain coefficients becomes difficult when the number of elements in the phased array shrinks. This is remedied through our approach by expanding the array excitations to include coefficients which correspond to real-valued phase rotations. Wherein our uniqueness is bolstered by our fidelity in quantizing the appropriate real number presenting the phase. This approach was shown to possess the relevant properties, in terms of uniqueness, while also providing the same benefits that are expected with DM. Namely, variability in the intensity of the radiation pattern through scrambling of the side lobes while preserving energy and symbol characteristics in the direction of the intended receiver. Networks of eavesdroppers with adequate spatial diversity who might attempt to decode transmitted messages are mitigated by distorting every symbol in a unique manner. Through this mechanism, a small amount of open communications security capacity is traded for vastly increased secrecy capacity against advanced threats.
\end{section}




\begin{thebibliography}{}
\providecommand{\url}[1]{#1}
\csname url@samestyle\endcsname
\providecommand{\newblock}{\relax}
\providecommand{\bibinfo}[2]{#2}
\providecommand{\BIBentrySTDinterwordspacing}{\spaceskip=0pt\relax}
\providecommand{\BIBentryALTinterwordstretchfactor}{4}
\providecommand{\BIBentryALTinterwordspacing}{\spaceskip=\fontdimen2\font plus
\BIBentryALTinterwordstretchfactor\fontdimen3\font minus
  \fontdimen4\font\relax}
\providecommand{\BIBforeignlanguage}[2]{{%
\expandafter\ifx\csname l@#1\endcsname\relax
\typeout{** WARNING: IEEEtran.bst: No hyphenation pattern has been}%
\typeout{** loaded for the language `#1'. Using the pattern for}%
\typeout{** the default language instead.}%
\else
\language=\csname l@#1\endcsname
\fi
#2}}
\providecommand{\BIBdecl}{\relax}
\BIBdecl

\end{thebibliography}


\begin{thebibliography}{1}
\providecommand{\url}[1]{#1}
\csname url@samestyle\endcsname
\providecommand{\newblock}{\relax}
\providecommand{\bibinfo}[2]{#2}
\providecommand{\BIBentrySTDinterwordspacing}{\spaceskip=0pt\relax}
\providecommand{\BIBentryALTinterwordstretchfactor}{4}
\providecommand{\BIBentryALTinterwordspacing}{\spaceskip=\fontdimen2\font plus
\BIBentryALTinterwordstretchfactor\fontdimen3\font minus
  \fontdimen4\font\relax}
\providecommand{\BIBforeignlanguage}[2]{{%
\expandafter\ifx\csname l@#1\endcsname\relax
\typeout{** WARNING: IEEEtran.bst: No hyphenation pattern has been}%
\typeout{** loaded for the language `#1'. Using the pattern for}%
\typeout{** the default language instead.}%
\else
\language=\csname l@#1\endcsname
\fi
#2}}
\providecommand{\BIBdecl}{\relax}
\BIBdecl

\bibitem{ortho_noise}
Y.~Ding and V.~Fusco, ``Orthogonal vector approach for synthesis of multi-beam
  directional modulation transmitters,'' \emph{IEEE Antennas and Wireless
  Propagation Letters}, vol.~14, pp. 1330--1333, 2015.

\end{thebibliography}


\begin{thebibliography}{1}
\bibliographystyle{IEEEtran}

\bibitem{DingFusco2014}
Y. Ding and V. F. Fusco, "A Vector Approach for the Analysis and Synthesis of Directional Modulation Transmitters," in {\it IEEE Transactions on Antennas and Propagation}, vol. 62, no. 1, pp. 361-370, Jan. 2014, doi: 10.1109/TAP.2013.2287001.

\bibitem{Stallings}
W. Stallings, ''Cryptography and Network Security.'' 4th ed. Pearson Education, India 2006.

\bibitem{DalyBernhard1}
M. P. Daly and J. T. Bernhard, “Beamsteering in pattern
reconfigurable arrays using directional modulation,”
{\it IEEE Trans. Antennas Propag.}, vol. 58, no. 7, pp. 2259-
2265, Jul., 2010.

\bibitem{DalyBernhard2}
M. P. Daly, E. L. Daly, and J. T. Bernhard,
“Demonstration of directional modulation using a
phased array,” IEEE Trans. Antennas Propag., vol. 58,
no. 5, pp. 1545-1550, May, 2010.

\bibitem{BlochBarros}
M. Bloch and J. Barros, ''Physical-Layer Security: From Information Theory to Security Engineering,'' Cambridge University Press, 2011.

\bibitem{DingFusco1}
Y. Ding and V. F. Fusco, "A review of directional modulation technology". {\it International Journal of Microwave and Wireless Technologies}, 8(7), 981-993, 2016. doi:10.1017/S1759078715001099

\bibitem{DingFusco2}
Y. Ding and V. F. Fusco, "Orthogonal Vector Approach for Synthesis of Multi-Beam Directional Modulation Transmitters," in {\it IEEE Antennas and Wireless Propagation Letters}, vol. 14, pp. 1330-1333, 2015, doi: 10.1109/LAWP.2015.2404818.

\bibitem{NegiGoel1}
R. Negi and S. Goel, ''Secret Communication Using Artificial Noise,'' in {\it Proc. IEEE Veh. Technol. Conf.}, Sept. 25-28 2005, vol. 62, no. 3, pp. 1906-1910.

\bibitem{NegiGoel2}
S. Goel and R. Negi, ''Guaranteeing Secrecy Using Artifical Noise,'' {\it IEEE Trans. Wireless Commun.}, vol. 7, no. 6, pp. 2180-2189, Jun., 2008.

\bibitem{Hafez2016}
M. Hafez, T. Khattab, T. Elfouly and H. Arslan, "Secure multiple-users transmission using multi-path directional modulation," {\it 2016 IEEE International Conference on Communications (ICC)}, Kuala Lumpur, Malaysia, 2016, pp. 1-5, doi: 10.1109/ICC.2016.7511289.

\bibitem{Fan2018}
X. Fan, Z. Zhang, W. Trappe, Y. Zhang, R. Howard and Z. Han, "Secret-Focus: A Practical Physical Layer Secret Communication System by Perturbing Focused Phases in Distributed Beamforming," {\it IEEE INFOCOM 2018 - IEEE Conference on Computer Communications}, Honolulu, HI, USA, 2018, pp. 1781-1789, doi: 10.1109/INFOCOM.2018.8486339.

\bibitem{DingFusco3}
Y. Ding and V. F. Fusco, "Directional modulation transmitter synthesis using particle swarm optimization," 2013 {\it Loughborough Antennas \& Propagation Conference (LAPC)}, Loughborough, UK, 2013, pp. 500-503, doi: 10.1109/LAPC.2013.6711950.

\bibitem{DingFusco4}
Y. Ding and V. F. Fusco, ''Establishing Metrics for Assessing the Performance of Directional Modulation Transmitters,'' 2014 {\it IEEE Transactions on Antennas and Propagation}, vol. 62, pp. 2745-2755

\bibitem{Shannon}
C.E. Shannon, ``Communication theory of secrecy systems,'' {\it The Bell Labs Technical Journal}, vol. 28, pp 656-715, 1949 doi:10.1002/j.1538-7305.1949.tb00928.x.

\bibitem{Wyner}
A.D. Wyner, ``The wire-tap channel,'' in {\it The Bell Labs Technical Journal}, vol. 54, no. 8, pp. 1355-1387, 1975 doi: 10.1002/j.1538-7305.1975.tb02040

\bibitem{BarrosRodr}
J. Barros and M. R. D. Rodrigues, ''Secrecy Capacity of Wireless Channels,'' {\it 2006 IEEE International Symposium on Information Theory}, Seattle, WA, 2006, pp. 356-360 doi: 10.1109/ISIT.2006.261613.

\bibitem{CsisKorn}
I. Csiszar and J. Korner, ''Broadcast Channels with Confidential Messages,'' {\it 1978 IEEE Transactions on Information Theory}, vol. 24, pp. 339-348.

\end{thebibliography}
\end{document}